\documentclass[aps,prc]{revtex4}
\usepackage{amssymb,amsmath,epsfig, color} 
\newcommand{\rf}[1]{(\ref{#1})}
\newcommand{\ba}{\begin{align}}
\newcommand{\ea}{\end{align}}
\newcommand{\lam}{\Lambda}

\newcommand{\be}{\begin{equation}}  
\newcommand{\ee}{\end{equation}}  
\newcommand{\bea}{\begin{eqnarray}}  
\newcommand{\eea}{\end{eqnarray}}  
\newcommand{\bmu}{\begin{multline}}
\newcommand{\emu}{\end{multline}}
\newcommand{\bean}{\begin{eqnarray*}}  
\newcommand{\eean}{\end{eqnarray*}}

\newcommand{\gapproxeq}{\lower  
.7ex\hbox{$\;\stackrel{\textstyle >}{\sim}\;$}}  
\newcommand{\lapproxeq}{\lower  
.7ex\hbox{$\;\stackrel{\textstyle <}{\sim}\;$}}

\newcommand{\bc}{\begin{center}}  
\newcommand{\ec}{\end{center}}  
\newcommand{\btab}{\begin{tabular}}  
\newcommand{\etab}{\end{tabular}}

\newcommand{\qq}{$q\overline q$}
\newcommand{\cc}{$c\overline c$}

\newcommand{\QQ}{$Q\overline Q$}

\newcommand{\ket}[1]{|           {#1}           \rangle}
\newcommand{\bra}[1]{\langle           {#1}           |}   

\newcommand{\tso}{^3\mathrm{S}_1}  
\newcommand{\grad}{\mathbf{\nabla}}

\newcommand{\rme}[3]{\bra{#1}|{#2}|\ket{#3}}
\newcommand{\me}[3]{\bra{#1}{#2}\ket{#3}}
\newcommand{\ot}{\otimes}

\newcommand{\uS}{\textrm{S}}
\newcommand{\uP}{\textrm{P}}
\newcommand{\uD}{\textrm{D}}
\newcommand{\uF}{\textrm{F}}

\newcommand{\epem}{e^+e^-}

\newcommand{\wb}[2]{[{#1}|{#2}]}

\newcommand{\wbk}[2]{[{#1}|{#2}]}

\def\S{\Sigma}
\def\P{\Pi}
\def\D{\Delta}

\def\bit{\begin{itemize}}
\def\eit{\end{itemize}}

\def\q{\overline q}
\def\Q{\overline Q}
\def\c{\overline c}

\def\PPm{1\P\uP^{-}}
\def\PPp{1\P\uP^{+}}

\begin{document}

\title{\bf Hadron production in $\psi$, $\eta_c$ and $\chi$ decays}

\author{
T. J. Burns\footnote{e-mail: burns@thphys.ox.ac.uk}
and F. E. Close\footnote{e-mail: F.Close1@physics.ox.ac.uk}
}
\affiliation{Department of Theoretical Physics,
University of Oxford, \\
Keble Rd., Oxford, OX1 3NP, United Kingdom}

\date{\today}

\begin{abstract}
We derive relations among branching fractions in the exclusive decay of charmonia to  light flavour %%@
meson pairs assuming 
factorization
between the quark spin and spatial degrees of freedom. With the further assumption that these 
amplitudes
can be described by flux-tube models, we assess prospects for production of hybrid mesons in 
charmonium decays.

\end{abstract}

\maketitle

%PACS numbers: 
%13.40.-f, 13.88.+e, 13.75.Jz
\section{Introduction}

The decays of $\psi$ to light hadrons via \qq~ intermediate OZI diagrams have been
investigated by several authors \cite{haber,seiden}. These analyses showed that while
leading OZI diagrams dominate and give a good first description of data, more
detailed fits implied that disconnected diagrams cannot be neglected in
general\cite{seiden}. These studies primarily related states of different flavour but 
having the same spin and spatial quantum numbers. To go beyond that it is neccessary to make an %%@
assumption concerning the intermediate \qq~ state. Even within the leading OZI
assumption, such efforts have foundered on the need to model the radial wavefunction of the
intermediate \qq~ state; the lack of knowledge of which radial excitations (`$n$')
dominate, and the need to model the specific wavefunctions, introduced model
dependences that severely limited useful predictions for the final states.

Recently we have shown \cite{bct} that results from lattice QCD imply a factorization of $L$ and $S$ %%@
in strong decays decays,
and that the created pair are spin triplet. This hypothesis 
factors out the spatial degrees of freedom and
gives relations (``L-S'' relations) among amplitudes for the decays of states sharing the same spatial %%@
quantum numbers but differing in spin and total angular momentum that are independent of
of the spatial wavefunctions.  
In this paper we apply such relations to the hadronic decays of charmonia to light flavour meson %%@
pairs, assuming that such decays can be modelled as the strong decay of an intermediate light quark %%@
\qq~ state.

By exploiting the factorization of amplitudes in OZI strong decays, we can avoid the $n$-dependence %%@
problem in so far as we leave the spatial part of the decay amplitude undetermined, and instead %%@
exploit the relations between decay amplitudes arising from the factorisation of L and S. Such %%@
relations apply in a symmetry limit, provided there is not a double conspiracy of a single $n$ %%@
dominating and kinematic node for that $n$. Within this more modest assumption we show that if the %%@
transition is indeed driven by a virtual \qq~ state, as implicitly assumed in analyses %%@
\cite{haber,seiden}, then some qualitative and semi-quantitative conclusions can be drawn about its %%@
nature and new relations among some final states obtained. 

We find that data rule out an intermediate \qq~ hybrid state and appear to be consistent with a %%@
coupling of \cc $\to$ \qq~ where the \qq~ is `canonical', in the sense of having the same $^{2S+1}L_J$ %%@
configuration as the initial \cc~ state. If confirmed by
further data this will provide a challenge for dynamical models to explain. It suggests that hybrid %%@
\qq~ production at short distances is
suppressed, which would imply that their production from electromagnetic or weak currents, e.g. in %%@
$e^+e^-$ or $B$ decays will be feeble. 

This leaves open the question of whether hybrids can be produced in charmonia decays in association %%@
with other hadrons. Within our model there is a selection rule suppressing the production of a %%@
negative parity hybrid along with a 1S meson; this applies to certain modes which $a$ $priori$ looked %%@
promising, such as  $\chi_1 \to \pi \pi_1$. On the other hand, we find that other exotic hybrids such %%@
as the $2^{+-}$ $b_2$ may be produced in sufficient measure to be observed, for instance in $\psi\to %%@
b_2\pi $.

In section \ref{model} we set up the model. In section \ref{ls} we confront the model predictions with %%@
data on charmonia decays. In section \ref{hybrids} we apply the model to the production of hybrid %%@
mesons.

\section{The model}\label{model}

\begin{figure}[ht]
%vqpc.tex
\setlength{\unitlength}{0.00023300in}%
\begingroup\makeatletter\ifx\SetFigFont\undefined
% extract first six characters in \fmtname
\def\x#1#2#3#4#5#6#7\relax{\def\x{#1#2#3#4#5#6}}%
\expandafter\x\fmtname xxxxxx\relax \def\y{splain}%
\ifx\x\y   % LaTeX or SliTeX?
\gdef\SetFigFont#1#2#3{%
  \ifnum #1<17\tiny\else \ifnum #1<20\small\else
  \ifnum #1<24\normalsize\else \ifnum #1<29\large\else
  \ifnum #1<34\Large\else \ifnum #1<41\LARGE\else
     \huge\fi\fi\fi\fi\fi\fi
  \csname #3\endcsname}%
\else
\gdef\SetFigFont#1#2#3{\begingroup
  \count@#1\relax \ifnum 25<\count@\count@25\fi
  \def\x{\endgroup\@setsize\SetFigFont{#2pt}}%
  \expandafter\x
    \csname \romannumeral\the\count@ pt\expandafter\endcsname
    \csname @\romannumeral\the\count@ pt\endcsname
  \csname #3\endcsname}%
\fi
\fi\endgroup
\begin{picture}(11713,8663)(5757,-16780)
\thicklines
\put(7951,-10936){\line(-1, 0){3300}}
\put(5101,-10936){\vector(-1, 0){2400}}
\put(5101,-10936){\line(-1, 0){3300}}
\put(10526,-10936){\vector(-1, 0){2000}}
\put(6151,-9961){\vector( 1, 0){3000}}
%\special{ps: gsave 0 0 0 setrgbcolor}
{\linethickness{1.4cm}
\put(3676,- 10586){\line(1,0){4550}}}
%\put(12751,-8161){\vector(-1,-1){1050}}
%\put(10951,-10936){\vector( 1,-1){975}}

%\put(13426,-12286){\vector(-1, 1){975}}
\put(12048,-10908){\vector(  1, -1){975}}

\put(13426,-8686){\vector(-1,-1){1050}}
\put(11626,-10486){\line( 1,-1){1800}}
\put(11626,-10486){\line( 1, 1){1800}}
\put(7651,-10936){\line( 1, 0){3300}}
\put(10951,-10936){\line( 1,-1){1800}}
\put(7651,-9961){\line( 1, 0){3300}}
\put(10951,-9961){\line( 1, 1){1800}}
\put(1801,-9961){\vector( 1, 0){1500}}
\put(5701,-10411){\makebox(6.6667,10.0000){\SetFigFont{10}{12}{rm}.}}
\put(1801,-9961){\vector( 1, 0){1500}}
\put(5701,-10411){\makebox(6.6667,10.0000){\SetFigFont{10}{12}{rm}.}}
\put(2401,-9961){\line( 1, 0){2700}}
\put(1801,-9961){\vector( 1, 0){1500}}
\put(5701,-10411){\makebox(6.6667,10.0000){\SetFigFont{10}{12}{rm}.}}
\put(1801,-9961){\vector( 1, 0){1500}}
\put(5701,-10411){\makebox(6.6667,10.0000){\SetFigFont{10}{12}{rm}.}}
%put(+6200,- 9500){$\bk{\vecR,\vecr}{(\lam\ot n'l')_{lm}}$}
%put(13500,-8500){$\bk{\vecr_2,\vecr_1}{(\lam_1\ot n_1'l_1')_{l_1m_1}}$}
%put(13700,-12500){$\bk{\vecr_1,\vecr_2}{(\lam_2\ot n_2'l_2')_{l_2m_2}}$}
%\put(9001,-12000){$\ket \bA$}
%\put(14001,- 8000){$\ket \bB$}
%\put(14001,-13000){$\ket \bC$}
\put(900,-11000){$\overline c$}
\put(900,-10000){$c$}
\put(13500,-8500){$\overline Q$}
\put(13000,-8000){$q$}
\put(13700,-12500){$Q$}
\put(13200,-13000){$\overline q$}
\end{picture}
\caption{\cc~ annihilation into a virtual \qq~ state which decays by the creation of a $Q\Q$ pair.}
\label{blackbox}
\end{figure}
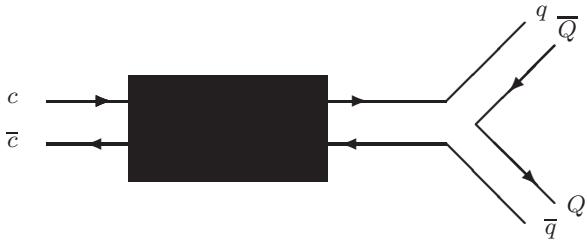

 We are interested in the decay topology in which the \cc~ annihilate into gluons or a virtual photon, %%@
creating a light quark $q\q$ pair which then decays by the creation of an additional light quark pair %%@
\QQ,
\be
c\c	\to 
q\q 	\to 
q\Q+Q\q,
\ee
as depicted in FIG. \ref{blackbox}.  Flavour disconnected decays $c\c\to q\bar{q}\to q\bar q+Q \bar Q$ %%@
are not negligible but are perturbations on the main phenomenology and will be ignored for this first %%@
survey.

We postulate that the intermediate $q\q$ state has well defined $j = l \oplus s$
and can be described as a non-relativistic quark-flux tube system, a  meson $m^{q\q}$, which decays by %%@
flux tube breaking in the usual way to mesons $m_{1}^{q\Q}+m_{2}^{Q\q}$. Thus we describe the %%@
transition
\be
m^{c\c}	\to
m^{q\q}	\to
m_{1}^{q\Q}+m_{2}^{Q\q}
\ee
by the amplitude
\be
\me{m_{1}^{q\Q}m_{2}^{Q\q}}{V}{m^{c\c}}
=\sum_{\ket{m^{q\q}_i}\bra{m^{q\q}_i}}
\me{m_{1}^{q\Q}m_{2}^{Q\q}}{H_I}{m^{q\q}_i}
\me{m^{q\q}_i}{V_{QCD}}{m^{c\c}}\label{amp1}
\ee
where 
\bit
\item $\ket{m^{q\q}_i}\bra{m^{q\q}_i}$ is a complete set of quark-flux tube states $m^{q\q}_i$ %%@
consistent with the $J^{PC}$ quantum numbers,
\item  $\me{m^{q\q}_i}{V_{QCD}}{m^{c\c}}$ is the amplitude for creation of the virtual state %%@
$m^{q\q}_i$ via $c\bar c \to ggg (\gamma^*) \to q\bar{q}$, and
\item  $\me{m_{1}^{q\Q}m_{2}^{Q\q}}{H_I}{m^{q\q}_i}$ is the amplitude for the decay of the virtual %%@
state $m^{q\q}_i$ via $q\q\to q\Q+Q\q$. 
\eit
Writing eqn. \rf{amp1} in another way, the full decay amplitude is proportional to the strong decay %%@
amplitude of an intermediate state  $\ket{m^{q\q}}$,
\be
\me{m_{1}^{q\Q}m_{2}^{Q\q}}{V}{m^{c\c}} = \me{m_{1}^{q\Q}m_{2}^{Q\q}}{H_I}{m^{q\q}}\label{amp2}
\ee
where $\ket{m^{q\q}}$ is a linear combination of possible configurations $m^{q\q}_i$ with mixing %%@
angles governed by the $V_{QCD}$ interaction,
\be
\ket{m^{q\q}}=\sum_{\ket{m^{q\q}_i}\bra{m^{q\q}_i}}
\ket{m^{q\q}_i}
\me{m^{q\q}_i}{V_{QCD}}{m^{c\c}}\label{mixed}
\ee
The complete set of intermediate virtual $q\q$ states $\ket{m^{q\q}_i}\bra{m^{q\q}_i}$ includes all %%@
possible orbital, radial and gluonic excitations consistent with the initial state quantum numbers, %%@
and $a$ $priori$ the hierarchy of mixing angles $\me{m^{q\q}_i}{V_{QCD}}{m^{c\c}}$ between such states %%@
is unknown. The amplitude $\me{m^{q\q}_i}{V_{QCD}}{m^{c\c}}$ describes a complicated non-perturbative %%@
process which we do not attempt to calculate; instead, we suppose that there is a unique choice of %%@
$m_i^{q\q}$ for which $\me{{m^{q\q}}_i}{V_{QCD}}{m^{c\c}}$ is large; that is to say, the wavefunction %%@
of $\ket{m^{q\q}}$ is dominated by a \qq~ state $\ket{m^{q\q}_i}$ in a unique
$S,L$ and $\Lambda$ configuration. With this assumption, the strong decay matrix element %%@
$\bra{m_{1}^{q\Q}m_{2}^{Q\q}}H_I\ket{{m^{q\q}}}$ relates the decay amplitudes of different final %%@
states using eqn. \rf{amp2}, subject only to the factorisation of spin and orbital angular momentum %%@
quantum numbers.

We will use the notation of Ref \cite{burns}. A quark-flux tube system can be described by a ket of %%@
the form
\be
\ket{(s\ot n\lam l^p)_j}
\ee
where  $s$, $l$ and $j$ are the usual spin, orbital and total angular momentum quantum numbers, $n$ is %%@
the quark radial quantum number, $\lam$ is the gluonic angular momentum and $p$ is a parity label %%@
appropriate to $\lam\ne 0$ states which occur in parity doublets. In anticipation of our proposal that %%@
the intermediate virtual state is dominated by a single configuration, let us associate %%@
$\ket{{m^{q\q}}}$ with a unique ket of the above form
\be
\ket{{m^{q\q}}}=\ket{(s\ot n\lam l^p)_j}.
\ee
The final state $\bra{m_{1}^{q\Q}m_{2}^{Q\q}}$ consists of a pair of mesons likewise defined, coupled %%@
to $j_{12}$  and in a 
relative partial wave $L$, in turn coupled to angular momentum $j$,
\be
\bra{m_{1}^{q\Q}m_{2}^{Q\q}}=\bra{(((s_1\ot n_1 \lam_1 l_1^{p_1})_{j_1}\ot(s_2\ot n_2 \lam_2 
l_2^{p_2})_{j_2})_{j_{12}}\ot L)_{jm}}.\label{finalstate}
\ee
In the flux tube model the decay operator $H_I$ is given by 
\be
H_I=\mathbf{\chi}_1\cdot\grad
\ee 
where $\chi_1$ creates a spin 1 $Q\Q$ pair and $\grad$ acts on the quark degrees of freedom. Thus we %%@
are interested in matrix elements of the form
\be
\bra{m_{1}^{q\Q}m_{2}^{Q\q}}H_I\ket{{m^{q\q}}}=\rme{(((s_1\ot n_1  \lam_1 l_1^{p_1})_{j_1}\ot(s_2\ot %%@
n_2 \lam_2 
l_2^{p_2})_{j_2})_{j_{12}}\ot 
L)_j}{\mathbf{\chi}_1\cdot\grad}{(s\ot n\lam l^p)_j}.
\label{fullme}
\ee
The decay rate is proportional to the sum over the squared transition amplitudes for all possible %%@
couplings $j_{12}$ and partial waves $L$; it is useful to define a corresponding bracket
\be
\wbk{m_{1}^{q\Q}m_{2}^{Q\q}}{{m^{q\q}}}=\sum_{L,j_{12}}
\rme{(((s_1\ot n_1  \lam_1 l_1^{p_1})_{j_1}\ot(s_2\ot n_2 \lam_2 
l_2^{p_2})_{j_2})_{j_{12}}\ot 
L)_j}{\mathbf{\chi}_1\cdot\grad}{(s\ot n\lam l^p)_j}^2.
\label{widthbracket}
\ee
Following eqn. \rf{amp2}, the branching fraction for a charmonium state to decay to a light meson pair %%@
$m_{1}^{q\Q}+m_{2}^{Q\q}$ is
\be
b.r.(
m^{c\c}	\to
m_{1}^{q\Q}+m_{2}^{Q\q})
\propto
pE_1E_2\wbk{m_{1}^{q\Q}m_{2}^{Q\q}}{m^{q\q}}\label{br}
\ee
where $E_1, E_2$ are the energies of mesons with mass $M_1,M_2$ produced with momentum $p$.

%The matrix element in eqn. \rf{fullme} has been calculated in the 
%$\tpn$ model for many different meson quantum numbers \cite{bcps} and generalised in the flux 
%tube model to the case of arbitrary quantum numbers and to include the possibility of outgoing 
%hybrid mesons\cite{burns}. As flux-tube models are the nearest simulation that we currently have 
%for the phenomenology that appears to occur in lattice QCD we adopt this approach here. The %model 
%successfully describes transitions among conventional mesons: $\S\to\S+\S$ \cite{ki}, and 
%also been applied to the decays of hybrid mesons, $\P\to\S+\S$ \cite{ikp,cp95}. In ref. 
%\cite{burns} these techniques have been extended to cover the {\it production} of a hybrid meson %in 
%the quasi two-body decays of a conventional meson: $\S\to\P+\S$ and $\S\to\P+\P$ . Thus for the 
%first time we have a formalism for predicting the relative production of hybrid and conventional 
%mesons in charmonium decays.

Our starting point will be the derivation of the matrix element
 \rf{fullme} as presented in refs. \cite{burns,tjbthesis}. In the approach presented there, the first %%@
step is to separate the spin and space degrees of freedom. This involves a recoupling of the states of %%@
good $j_1,j_2,j_{12}$ into states of good $s_{12},l_{12},l_f$, whence the spin matrix element is %%@
simply given in terms of a 9-$j$ coefficient. In this way the full matrix element
\be
\rme{(((s_1\ot n_1  \lam_1 l_1^{p_1})_{j_1}\ot(s_2\ot n_2 \lam_2 
l_2^{p_2})_{j_2})_{j_{12}}\ot 
L)_j}{\mathbf{\chi}_1\cdot\grad}{(s\ot n\lam l^p)_j},
\ee
is expressed as linear combination of spatial matrix elements of the form
\be
\rme{((n_1\lam_1 l_1^{p_1}\ot n_2\lam_2 l_2^{p_2})_{l_{12}}\ot L)_{l_f}}{\grad}{n\lam l^p}.
\label{spaceme}
\ee
This leads to relationships among amplitudes to different final states sharing the same spatial %%@
quantum numbers but differing
in spin and total angular momentum. Such relations apply in the limit that the final states under %%@
comparison have the same masses, radial wavefunctions and decay momenta. The relations are also %%@
independent of the radial wavefunction of the initial state, hence they translate directly into %%@
relations among the branching fractions using eqn. \rf{br} provided that there is not a double %%@
conspiracy in which $ m^{q\q}$ is dominated by a single radial configuration $n$ which has a kinematic %%@
node suppressing certain channels.

\section{L-S relations among amplitudes}\label{ls}
	
We now turn to the question of the intermediate state $\ket{m^{q\q}}$, which can in general be a %%@
linear combination of states  $\ket{m^{q\q}_i}$ consistent with the $J^{PC}$ quantum numbers of %%@
$\ket{m^{c\c}}$, as in eqn \rf{mixed}. For the charmonia states of interest, the angular momentum %%@
configurations consistent with the  $J^{PC}$ quantum numbers are \cite{tjbthesis}:
\bea
m^{c\c}	&	& m^{q\q}_i\nonumber\\
\eta_c	&\qquad&^1\S\uS_0,^3\P\uP_0	\label{laterref1}\\
\psi	&\qquad&^3\S\uS_1,^3\S\uD_1,^1\P\uP_1,^3\D\uD_1\\
%1^{+-}	&\qquad&^1\S\uP_1,^3\P\uP_1,^3\P\uD_1	\\
\chi_0	&\qquad&^3\S\uP_0          	\\
\chi_1	&\qquad&^3\S\uP_1,^1\P\uP_1,^3\D\uD_1\\
\chi_2	&\qquad&^3\S\uP_2,^3\S\uF_2,^1\P\uD_2,^3\D\uD_2,^3\D\uF_2\label{laterref2}
\eea
where in the above we have used the parlance of molecular physics, in which the quantum numbers %%@
$l=0,1,2,\ldots$ are labelled S,P,D\ldots and   $\lam=0,1, 2\ldots$ are analogously labelled %%@
$\Sigma,\Pi,\Delta\ldots$.

In the following sections we  discriminate among the possible angular momentum configurations $s$, %%@
$\lam$ and $l$ for the intermediate states   $ m^{q\q}_i$ with reference to data on $\eta_c$, $\chi$ %%@
and $\psi$ decays. In ref. \cite{tjbthesis} it was shown that the selection rule of refs. %%@
\cite{ikp,cp95} forbidding the decay of  $\Pi\uP$ hybrids to identical S-wave mesons applies also to %%@
$\P\uD$ and $\Delta\uD$-type hybrids. Using this, we can rule out hybrid configurations in $\eta_c$ %%@
and $\chi$ decays which show prominent production of identical S-wave pairs. In the case of $\psi$ the %%@
spin singlet selection rule argues against the hybrid intermediate state. 

We hypothesise that $ m^{q\q}$ is dominated by the $\S$-state with the smallest $l$ consistent with %%@
the quantum numbers; we refer to this as the ``canonical choice'' wavefunction. Within this hypothesis %%@
we express the production rates as linear combinations of spatial matrix elements of the type %%@
\rf{spaceme} with the superfluous $\lam$ and $p$ labels dropped,
\be
\rme{((n_1l_1\ot n_2 l_2)_{l_{12}}\ot L)_{l_f}}{\grad}{n l}.
\label{spacemereduced}
\ee
The resulting L-S relations are consistent with data and make predictions for modes which have not yet %%@
been observed.

\subsection{$\eta_c$ decays}

The virtual $q\q$ state $m^{q\q}$ corresponding to the $\eta_c$ could have $^1\S\uS_0$ or $^3\P\uP_0$ %%@
assignments.  On heuristic grounds we expect the latter to be suppressed owing to a node in both the %%@
quark and flux tube wavefunctions at the origin. If the virtual $q\q$ corresponding to the $\eta_c$ %%@
were dominantly $^3\P\uP_0$, the production of identical $^3\uS_1$ pairs would be forbidden not only %%@
by the standard hybrid selection rule forbidden the decay to identical S-wave pairs \cite{ikp, cp95} %%@
but also by a more general selection 
rule derived in ref. \cite{tjbthesis}, 
\be
\wb{^3\uS_1~^3\uS_1}{^3\P\uP^-_0}=0,
\ee
which relies only on L-S factorisation. In this context it is notable that three vector-vector modes %%@
have been observed with considerable branching fractions \cite{pdg},
\bea
b.r.(\eta_c  \to	\rho\rho	)&=&(20\pm 7) 	\times 10^{-3}\\
b.r.(\eta_c  \to	K^*\bar{K^*}		)&=&(9.2\pm 3.4) 	\times 10^{-3}\\
b.r.(\eta_c  \to	\phi\phi	)&=&(2.7\pm 0.9) 	\times 10^{-3}
\eea
These branching fractions are comparable with multi-body modes
\bea
b.r.(\eta_c \to	\eta'\pi\pi				)&=&(41	\pm 7  ) 	\times 10^{-3}\\
b.r.(\eta_c \to	K^{*0}K^-\pi^+ + c.c.			)&=&(20 	\pm 7  ) 	\times 10^{-3}\\
b.r.(\eta_c \to	K^{*0}\overline{K^{*0}}\pi^+\pi^- 	)&=&(15 	\pm 8  ) 	\times 10^{-3}\\
b.r.(\eta_c \to	\phi K^+ K^-				)&=&(2.9	\pm 1.4) 	\times 10^{-3}
\eea
and for $\rho\rho$, at least, exceed those of the 1P+1S modes expected to dominate for decay of  %%@
$^3\P\uP_0$ state,
\bea
b.r.(\eta_c \to	a_2\pi					)&<& 20	 	\times 10^{-3}\\
b.r.(\eta_c \to	f_2\eta               			)&<& 11		\times 10^{-3}
\eea
This supports the conjecture that  $m^{q\q}$ will be dominated by the $^1\S\uS_0$ rather than %%@
$^3\P\uP_0$ 
configuration. Assuming therefore the canonical choice $^1\S\uS_0$ wavefunction dominates,  one can %%@
express the allowed decay amplitudes in terms of common spatial matrix elements and therefore extract %%@
relations between them. For decays to S-wave pairs, only vector-vector pairs are allowed. For P+S %%@
final states there are several possible decays, each of which can be expressed uniquely in terms of S- %%@
and D-wave spatial matrix elements $S_1$ and $D_1$ defined
\bea
S_1	&=&\rme{((n_1\uP\ot n_2\uS)_1\ot \uS)_1}{\grad}{n\uS}\label{S1}\\
D_1	&=&\rme{((n_1\uP\ot n_2\uS)_1\ot \uD)_1}{\grad}{n\uS}\label{D1}
\eea
In the above we leave the radial quantum numbers $n$, $n_1$ and $n_2$ explicit as a reminder of the %%@
fact that the relations we derive are independent of the radial wavefunctions of the initial and final %%@
states. If specific models are subsequently developed, the resulting
wavefunctions can be applied to these immediately. The appropriate decompositions are derived in ref. %%@
\cite{burns} and are presented explicitly in ref. \cite{tjbthesis}. For the width bracket of the form %%@
\rf{widthbracket}, the result is
\bea
\wb{^3\uP_0~ ^1\uS_0 }{^1\uS_0}&=	&\frac{{{S_1}}^2}{4}\\
\wb{^3\uP_2~ ^1\uS_0 }{^1\uS_0}&=	&\frac{{{D_1}}^2}{4}\\
\wb{^1\uP_1~ ^3\uS_1 }{^1\uS_0}&=	&\frac{{{S_1}}^2 + {{D_1}}^2}{4}
\eea
and thus
\be
\wb{^3    \uP_0~^1\uS_0}{^1\uS_0}+\wb{^3\uP_2~^1\uS_0}{^1\uS_0}=\wb{^1\uP_1
~ ^3\uS_1}{^1\uS_0}
\ee
Thus the decay rates for isovector modes are related
\be
\wb{   a_0\pi   }{\eta_c}+\wb{   a_2\pi   }{\eta_c}=\wb{b_1\rho     }{\eta_c}\label{iv}
\ee
At present there is an upper limit on the $a_2\pi$ mode \cite{pdg}; if this mode can be isolated then %%@
eqn. \rf{iv} relates the branching fractions of $b_1\rho$ and $a_0\pi$ and this prediction could be %%@
confronted with experiment. For isoscalar pairs the 
analogous relationships follow immediately from the above with appropriate mixing angles. For mixed %%@
$K_1$-$K_1'$ states 
\bea
\ket{K_1}&=&\cos\phi\ket{^1\S\uP_1}+\sin\phi\ket{^3\S\uP_1}\\
\ket{K_1'}&=&-\sin\phi\ket{^1\S\uP_1}+\cos\phi\ket{^3\S\uP_1}.
\eea
only the spin singlet part contributes to the decay amplitude, hence we predict
\bea
\wb{K_0 K }{ \eta_c}+\wb{K_2 K }{ \eta_c}&=&\wb{K_1 K^*}{\eta_c}+\wb{K_1'K^*}{\eta_c}.
\eea

\subsection{$\chi_0$ decays}

For $\chi_0$ decays, the only quantum number assignment available for $m^{q\q}$ is $^3\S\uP_0$. %%@
Proceeding as before, we use the decompositions presented in ref. \cite{tjbthesis} to express the %%@
allowed decay modes in terms of common spatial matrix elements. Both pseudoscalar and vector pairs are %%@
allowed, and the corresponding amplitudes are expressed in terms of two spatial matrix elements $S$ %%@
and $D$, defined
\bea
S	&=&\rme{((n_1\uS\ot n_2\uS)_0\ot \uS)_0}{\grad}{n\uP}\label{S}\\
D	&=&\rme{((n_1\uS\ot n_2\uS)_0\ot \uD)_2}{\grad}{n\uP}\label{D}
\eea
The decompositions are
\bea
\wb{^1\uS_0~ ^1\uS_0 }{^3\uP_0}&=	&\frac{{{S}}^2}{4}\\
\wb{^3\uS_1~ ^3\uS_1 }{^3\uP_0}&=	&\frac{{{S}}^2 + 4\,{{D}}^2}{12}
\eea
from which we obtain the constraint
\be
\wb{^3\uS_1~^3\uS_1}{^3\uP_0} \geq 1/3\wb{^1\uS_0~^1\uS_0}{^3\uP_0}
\ee
There are large phase space and momenta differences for $^3\uS_1~^3\uS_1$ and 
$^1\uS_0~^1\uS_0$ modes, and to quantify these would require
specific model wavefunctions. The experimental modes \cite{pdg}
\bea
b.r.(\chi_0\to K^{*0}\overline{K^{*0}})	&=&1.8\pm 0.6 \times 10^{-3}\\
b.r.(\chi_0\to K^+K^-)			&=&5.4\pm 0.6 \times 10^{-3}
\eea
and also the $\omega \omega; \eta \eta$ and $\pi\pi$ are nonetheless consistent with the above. 

Decays to P+S final states can be expressed in terms of a spatial matrix element $P_1$ defined as
\be
P_1	=\rme{((n_1\uP\ot n_2\uS)_1\ot \uP)_1}{\grad}{n\uP}\label{P1}
\ee
and the decompositions are
\be
\wb{^3\uP_1~ ^1\uS_0 }{^3\uP_0}=
\wb{^1\uP_1~ ^3\uS_1 }{^3\uP_0}=\frac{{{P_1}}^2}{6}.
\ee
Thus we predict
\be
\wb{a_1\pi}{\chi_0}=\wb{ b_1\rho}{\chi_0}
\ee
and likewise for strange pairs
\be
\wb{K_1K}{\chi_0}+\wb{K_1'K}{\chi_0}=\wb{K_1K^*}{\chi_0}+\wb{K_1'K^*}{\chi_0}
\ee
These will be a useful experimental test of our hypothesis.

\subsection{$\chi_1$ decays}

For the $1^{++}$ sector, the $m^{q\q}$ assignments could be the canonical $^3\S\uP_1$, or hybrid 
configurations
$^1\P\uP_1,^3\D\uD_1$. 
The  first two have the same penalty for the quark wavefunction at the origin, but it 
is anticipated that the $^3\S\uP_1$ will win over the $^1\P\uP_1$ because the latter 
will also have a penalty for the creation of a flux tube at the origin: it is not possible
 to have a transverse excitation in a flux tube with zero size. 
 
 The only two-body mode that
  has been directly observed is $K^*K^*$: that this has comparable strength to multi-body 
  modes argues against the $^1\P\uP_1,^3\D\uD_1$ assignments since the production of these identical %%@
particles would be forbidden owing to the (generalised) hybrid selection rule of ref. %%@
\cite{tjbthesis}. Hence the hybrid intermediate states are ruled out and we consider the L-S relations %%@
among P+S modes assuming the canonical $^3\S\uP_1$ dominates. The allowed decay amplitudes are %%@
expressed in terms of the spatial matrix element $P_1$ defined in eqn. \rf{P1} and three further %%@
spatial matrix elements $P_0$, $P_2$ and $F_2$,
\bea
P_0	&=&\rme{((n_1\uP\ot n_2\uS)_1\ot \uP)_0}{\grad}{n\uP}\label{P0}\\
P_2	&=&\rme{((n_1\uP\ot n_2\uS)_1\ot \uP)_2}{\grad}{n\uP}\label{P2}\\
F_2	&=&\rme{((n_1\uP\ot n_2\uS)_1\ot \uF)_2}{\grad}{n\uP}\label{F2}
\eea
The decompositions are
\bea
\wb{^3\uP_0~ ^1\uS_0 }{^3\uP_1}&=	&\frac{4\,{{P_0}}^2 + 3\,{{P_1}}^2 + 2\,{\sqrt{15}}\,{P_1}\,{P_2} %%@
+ 5\,{{P_2}}^2 - 4\,{P_0}\,\left( {\sqrt{3}}\,{P_1} + {\sqrt{5}}\,{P_2} \right) }{216}\\
\wb{^3\uP_1~ ^1\uS_0 }{^3\uP_1}&=	&\frac{16\,{{P_0}}^2 + 3\,{{P_1}}^2 - 2\,{\sqrt{15}}\,{P_1}\,{P_2} %%@
+ 5\,{{P_2}}^2 + {P_0}\,\left( -8\,{\sqrt{3}}\,{P_1} + 8\,{\sqrt{5}}\,{P_2} \right) }{288}\\
\wb{^3\uP_2~ ^1\uS_0 }{^3\uP_1}&=	&\frac{80\,{{P_0}}^2 + 15\,{{P_1}}^2 - %%@
2\,{\sqrt{15}}\,{P_1}\,{P_2} + {{P_2}}^2 + 8\,{P_0}\,\left( 5\,{\sqrt{3}}\,{P_1} - {\sqrt{5}}\,{P_2} %%@
\right)  + 36\,{{F_2}}^2}{864}\\
\wb{^1\uP_1~ ^3\uS_1 }{^3\uP_1}&=	&\frac{4\,{{P_0}}^2 + {{P_1}}^2 + {{P_2}}^2 + {{F_2}}^2}{24}
\eea
from which we obtain the relation
\be
\wb{^1\uP_1~^3\uS_1~}{^3\uP_1}=\wb{^3\uP_0 ~ %%@
^1\uS_0}{^3\uP_1}+\wb{^3\uP_1~^1\uS_0}{^3\uP_1}+\wb{^3\uP_2~^1\uS_0}{^3\uP_1}
\ee
This implies
\be
\wb{\rho b_1}{\chi_1}=\wb{\pi a_0}{\chi_1}+\wb{\pi a_1}{\chi_1}+\wb{\pi a_2}{\chi_1}
\ee
and
\be
\wb{K^*K_1}{\chi_1}+\wb{K^*K_1'}{\chi_1}
=
\wb{KK_0}{\chi_1}+\wb{KK_1}{\chi_1}+\wb{KK_1'}{\chi_1}+\wb{KK_2}{\chi_1}
\ee
Data on $\chi_1$ decays are sparse and we urge that tests of the above relations be investigated.

\subsection{$\chi_2$ decays}

For $\chi_2$ decays there are many possible assignments for the virtual \qq~ state: 
two conventional states $^3\S\uP_2,^3\S\uF_2$;
and three hybrids $^1\P\uD_2,^3\D\uD_2,^3\D\uF_2$. The empirical prevalence of \cite{pdg} 
\be
\chi_2\to\phi\phi,\omega\omega,\pi\pi,\eta\eta
\ee
once again rules out hybrid interpretations. The two $\S$ configurations can be discriminated by %%@
considering the L-S relations between pseudoscalar and vector pair amplitudes. For $^3\uP_2$ the %%@
decays can be expressed in terms of the spatial matrix elements $S$ and $D$ of eqns. \rf{S} and %%@
\rf{D},
\bea
\wb{^1\uS_0~ ^1\uS_0 }{^3\uP_2}&=	&\frac{{{D}}^2}{20}\\
\wb{^3\uS_1~ ^3\uS_1 }{^3\uP_2}&=	&\frac{5\,{{S}}^2 + 2\,{{D}}^2}{15}
\eea
and hence there is a constraint
\be
\wb{^3\uS_1~^3\uS_1}{^3\uP_2}> \frac{8}{3}\wb{^1\uS_0~^1\uS_0}{^3\uP_2}.
\ee
By contrast, for $^3\uF_2$ one obtains 
\be
\wb{^3\uS_1~^3\uS_1}{^3\uF_2}= \frac{11}{21}\wb{^1\uS_0~^1\uS_0}{^3\uF_2}.
\ee
Experimentally \cite{pdg}
\bea
b.r.(\chi_2\to K^{*0}\overline{K^{*0}})	&=&(38 \pm 9)\times 10^{-4}\\
b.r.(\chi_2\to K^+ K^-)						&=&(7.7 \pm 1.4)\times 10^{-4}\\
b.r.(\chi_2\to K_sK_s)						&=&(6.7\pm 1.1)        \times 10^{-4}
\eea
which rules out the $^3\uF_2$ and once again the canonical configuration, in this 
case $^3\uP_2$, dominates. For isovectors, if we associate the $2(\pi^+\pi^-)$ mode with $\rho\rho$, %%@
the experimental result concurs:
\be
b.r.(\chi_2\to \rho\rho)\approx 6\times b.r.(\chi_2\to \pi\pi)
\ee
Amplitudes for the P+S modes are defined in terms of the spatial matrix elements $P_1$, $P_2$ and %%@
$F_2$ of eqns. \rf{P1}, \rf{P2} and \rf{F2}, and the decompositions are
\bea
\wb{^3\uP_1~ ^1\uS_0 }{^3\uP_2}&=	&\frac{5\,{{P_1}}^2 - 6\,{\sqrt{15}}\,{P_1}\,{P_2} + 27\,{{P_2}}^2 %%@
+ 12\,{{F_2}}^2}{480}\\
\wb{^3\uP_2~ ^1\uS_0 }{^3\uP_2}&=	&\frac{5\,{{P_1}}^2 + 2\,{\sqrt{15}}\,{P_1}\,{P_2} + 3\,{{P_2}}^2 %%@
+ 8\,{{F_2}}^2}{160}\\
\wb{^1\uP_1~ ^3\uS_1 }{^3\uP_2}&=	&\frac{5\,{{P_1}}^2 + 9\,\left( {{P_2}}^2 + {{F_2}}^2 \right) %%@
}{120}
\eea
This yields a relation
\be
\wb{^3\uP_1~^1\uS_0}{^3\uP_2}+\wb{^3\uP_2~^1\uS_0}{^3\uP_2}=\wb{^1\uP_1~^3\uS_1}{^3\uP_2}
\ee
which can be tested experimentally.

 \subsection{$\psi$ decays}

For $\psi$ the possible
intermediate states $m_i^{q\q}$ are ${^3\Sigma \uS_1},  {^3\Sigma \uD_1},{^1\Pi\uP_1},{^3\D\uD_1}$.  
That the \qq~ pair is created at a point suggests ${^3\Sigma \uS_1}$ will dominate, and 
this appears to be confirmed by the data: any significant $\Pi $ or $\D$ admixture would be 
in conflict with known prevalence of experimental modes such as
\bea
\epem&\to&\eta_c\psi\textrm{, and}\\
\psi &\to&\rho \pi 
\eea
owing to the standard selection rule, although this rule may be 
broken quite significantly for final states with different spatial wavefunctions
\cite{closedudek}. A stronger rule appears assuming only the factorisation of L and S in the form of %%@
the spin-singlet selection rule, which forbids decays of the type
\be
\textrm{spin 0} \to \textrm{spin 0} + \textrm{spin 0}.
\ee
Experimentally, the decay
\be
\psi\to b_1\pi	
\ee
is large; this would be forbidden if the virtual \qq~ is in the hybrid $^1\Pi\uP_1$ state. 

With the canonical choice of $\tso$ dominance we can express the decay rates to P+S states in terms of %%@
the spatial matrix elements $S_1$ and $D_1$ of eqns. \rf{S1} and \rf{D1},
\bea
\wb{^1\uP_1~ ^1\uS_0 }{^3\uS_1}&=	&\frac{{{S_1}}^2 + {{D_1}}^2}{12}\\
\wb{^3\uP_0~ ^3\uS_1 }{^3\uS_1}&=	&\frac{{{S_1}}^2}{4}\\
\wb{^3\uP_1~ ^3\uS_1 }{^3\uS_1}&=	&\frac{4\,{{S_1}}^2 + {{D_1}}^2}{12}\\
\wb{^3\uP_2~ ^3\uS_1 }{^3\uS_1}&=	&\frac{{{D_1}}^2}{2}\label{earlier}\\
\eea
and hence there are two independent relations
\bea
\wb{^1\uP_1~^1\uS_0}{^3\uS_1}&=&\frac{1}{3}\wb{^3\uP_0~^3\uS_1}{^3\uS_1}+\frac{1}{6}\wb{^3\uP_2~^3\uS_%%@
%%@
1}{^3\uS_1}
\label{gg1}  \\
\wb{^3\uP_1~^3\uS_1}{^3\uS_1}&=&\frac{4}{3}\wb{^3\uP_0~^3\uS_1}{^3\uS_1}+\frac{1}{6}\wb{^3\uP_2~^3\uS_%%@
%%@
1}{^3\uS_1}
\label{gg2}
\label{psirelations}
\eea
Thus for $\psi$ decays to isovector pairs the model predicts
\bea
\wb{b_1\pi}{\psi}&=&\frac{1}{3}\wb{a_0\rho}{\psi}+\frac{1}{6}\wb{a_2\rho}{\psi}\label{qw1}\\
\wb{a_1\rho}{\psi}&=&\frac{4}{3}\wb{a_0\rho}{\psi}+\frac{1}{6}\wb{a_2\rho}{\psi}\label{qw2}
\eea
Experimentally the modes $a_2\rho$ and $b_1\pi$ have been measured
\bea
b.r.(\psi\to b_1\pi)&=&(5.3 \pm 0.8)\times 10^{-3}\\
b.r.(\psi\to a_2\rho)&=&(10.9 \pm 2.2)\times 10^{-3}
\eea
and are in agreement with the weak constraint from the above
\be
\wb{b_1\pi}{\psi}\geq\frac{1}{6}\wb{a_2\rho}{\psi}.
\ee
Using eqns. \rf{qw1} and \rf{qw2} gives a prediction for the scale of the as yet unseen $a_0\rho$ and 
$a_1\rho$ modes. As a first estimate we approximate the experimental result by %%@
$\wb{b_1\pi}{\psi}=\frac{1}{2}\wb{a_2}{\rho}$
and thus from eqns \rf{qw1} and \rf{qw2}
\bea
\wb{a_0\rho}{\psi}&=&2\wb{b_1\pi}{\psi}\\
\wb{a_1\rho}{\psi}&=&3\wb{b_1\pi}{\psi}
\eea
hence we predict
\bea
b.r.(\psi\to a_0\rho)&\approx 	&10 \times 10^{-3}\\
b.r.(\psi\to a_1\rho)&\approx	&15 \times 10^{-3}
\eea
These modes will be difficult to identify but feed multi-body channels which are observed to be 
large,
\bea
a_1\rho\to(\rho\pi)\rho&\to&5\pi\\
a_0\rho\to(\omega\pi\pi)\rho&\to&\omega 4\pi
\eea
The $\omega\pi^+\pi^+\pi^-\pi^-$ mode is observed with branching fraction $(8.5 \pm 3.4)\times %%@
10^{-3}$ which is consistent with the above prediction.

For strange pairs are the analogues of eqns \rf{qw1} and \rf{qw2} are
\bea
\wb{K_1K}{\psi}+\wb{K_1'K}{\psi}&=&\frac{1}{3}\wb{K_0K^*}{\psi}+\frac{1}{6}\wb{K_2K^*}{\psi}\label{ggg%%@
}\\
\wb{K_1K^*}{\psi}+\wb{K_1'K^*}{\psi}&=&\frac{4}{3}\wb{K_0K^*}{\psi}+\frac{1}{6}\wb{K_2K^*}{\psi}
\eea
Experimentally
\bea
b.r.(\psi\to K^{*0}\overline{K_2^0}+c.c.)	&=&(6.7 \pm 2.6)\times 10^{-3}\\
b.r.(\psi\to K_1(1400)^\pm K^\mp)		&=&(3.8 \pm 1.4)\times 10^{-3}\\
b.r.(\psi\to K_1(1270)^\pm K^\mp)		&<& 3.0         \times 10^{-3}
\eea
If the $K_1(1270)^\pm K^\mp$ mode can be measured then the $K_0K^*$ branching fraction is predicted by %%@
eqn. \rf{ggg}.

Analysis of the ${^3\Sigma \uD_1}$ possibility is in general model dependent. There
is however a potential discriminator in the helicity selection rule of ref\cite{bct},
which states that for ${^3\Sigma \uS_1}$ the amplitude for decay to vector and tensor
\qq~, such as $\omega f_2$, vanishes when the tensor has helicity $\pm 2$. For 
${^3\Sigma \uD_1}$ by contrast, this amplitude is not suppressed.

\section{Hybrid meson production}\label{hybrids}

In this section we consider the prospects for hybrid production in charmonia decay. We take as our %%@
starting point the hypothesis confirmed in the previous section that the intermediate state $m^{q\q}$ %%@
is dominated by a conventional $\S$ state. With this assumption we have expressed decay rates to %%@
conventional meson final states with spatial quantum numbers $n_1l_1$ and $n_2l_2$ as linear %%@
combinations of spatial amplitudes of the form
\be
\rme{((n_1 l_1\ot n_2 l_2)_{l_{12}}\ot L)_{l_f}}{\grad}{n l}.
\label{sme1}
\ee
Since these amplitudes are in general independent it is not possible to relate branching fractions %%@
among states with different spatial quantum numbers $n_1l_1$ and $n_2l_2$. 

Following the same approach, the decay amplitude to final states with spatial quantum numbers %%@
$n_1\lam_1 l_1^{p_1}$ and  $n_2\lam_2 l_2^{p_2}$ can be expressed in terms of spatial amplitudes of %%@
the form  
\be
\rme{((n_1\lam_1 l_1^{p_1}\ot n_2\lam_2 l_2^{p_2})_{l_{12}}\ot L)_{l_f}}{\grad}{n l}.
\label{sme2}
\ee
The novel feature of the approach presented in refs. \cite{burns,tjbthesis} is that all spatial %%@
amplitudes of the form \rf{sme2} are linearly related to spatial amplitudes of the form \rf{sme1}, %%@
leading in some cases to direct relationships between hybrid production amplitudes and those of %%@
conventional mesons. 

With this approach, it is possible to make general statements about hybrid production rates without %%@
having to make further assumption concerning the virtual state $m^{q\q}$. We restrict the discussion %%@
here to the production of a $1\P\uP^\pm$ hybrid along with a 1S meson, for which the linear relations %%@
between matrix elements \rf{sme2} and \rf{sme1} are particularly simple:
\bea
\rme{((1\P\uP^-\ot 1\uS)_\uP\ot L)_{l_f}}{\grad}{nl}&=&0\label{rel1}\\
\rme{((1\P\uP^+\ot 1\uS)_\uP\ot L)_{l_f}}{\grad}{n l}
			&\approx&\sqrt\frac{1}{2}\rme{((1\uS\ot 1\uP)_\uP\ot L)_{l_f}}{\grad}{nl}.\label{rel2}
\eea
The expression \rf{rel1} is a selection rule forbidding the decay of any $\S$ state to a $1\P\uP^-$ %%@
hybrid in the limit that the hybrid quark radial wavefunction is a 1P harmonic oscillator with the %%@
same size as the recoiling 1S meson. The second expression contains a ''$\approx$'' sign to indicate %%@
that it is calculated with a first order expansion in the hybrid radial wavefunction. Further details %%@
of these approximations can be found in refs \cite{burns,tjbthesis}.

In light of the experimental candidates  $\pi_1(1600),\pi_1(1400)$, the production of 
the exotic $1^{-+}$ along with a 
1S meson is an interesting possibility. A priori the most favourable production modes from 
$\psi$ and $\chi$ states might be expected to have been
\bea
\psi		&\to	&^3\PPm_1 ~1^3\uS_1	\label{mode1}\\
\chi_{0,1,2}	&\to	&^3\PPm_1~1^1\uS_0 	\label{mode2}
\eea
In the proposed model the decay proceeds by the strong decay of a virtual state $m^{q\q}$ in a $\S$ %%@
configuration. In this case the above modes should be strongly suppressed owing 
to a selection rule \rf{rel1} above, as should the analogous production modes of the non-exotic %%@
$J^{PC}$ hybrids $1^3\Pi\uP^-_{0,2}$ 
and $1^1\Pi\uP^-_1$ belonging to the same family. Thus, for instance, observation of significant modes
\bea
\psi&\to& \pi_1\rho\label{psitopi1rho}\\
\chi_1&\to& \pi_1\pi\label{chitopi1pi}
\eea
in the $\pi_1(1400),\pi_1(1600)$ channels would argue against a $\PPm$ hybrid interpretation for 
them (see e.g. ref \cite{burns}). 

For charmonia states the non-relativistic approximation should be more robust, but there are 
questions as to whether the dynamics 
at this scale are driven by flux tube breaking. In general the selection rule presented here will 
be broken if the decay mechanism 
is dominated by perturbative gluons. On mass grounds it is possible that the X(3940) observed in
\be
\epem\to \psi X(3940)\label{x3940},
\ee
contains a hybrid state with $J^{-+}$ or $J^{++}$ quantum numbers. There has also been suggestion 
that the $1^{--}$ $Y(4260)$ is a $1^1\Pi\uP^-_1$ hybrid \cite{cp95}: the 
immediate implication is that it should have $(0,1,2)^{-+}$ partners $1^3\Pi\uP^-_{0,1,2}$, and 
notably the $0^{-+}$ and  $1^{-+}$ 
are expected to be lighter \cite{rumsfeld}. 
If $\epem\to c\c+c\c$ can be modelled in a way analogous to the model presented here, a hybrid with %%@
$J^{-+}$ 
quantum numbers is not expected to be produced in the above reaction; however if the dominant %%@
production is via
gluon exchange \cite{braaten,bct} such hybrids could be produced. Predictions for such
a mechanism require models that go beyond the present discussion.

Subject to the proviso that production is by strong OZI, flux-tube breaking, rather than single 
gluon exchange, the selection rule may help discriminate between hybrid and other exotic or 
non-exotic interpretations of heavy or light-quark 
states, including those with non-exotic quantum numbers. 
%Experimentally 
%\be
%\epem\to \psi\eta_c
%\ee
%is observed to be large \cite{ex3940}, and so in a qualitative sense the production of other 
%S-wave pairs may also be expected to 
%be large. If the mode
%\be
%\epem \to Y(4260)\eta_c.
%\ee
%could be isolated this would support the interpretation of the $Y(4260)$ as a conventional 
%radially excited 1S state. If flux tube 
%formation and breaking is dominant then the above mode will be suppressed for a hybrid 
%$Y(4260)$.

The picture is altogether different for the production of the positive parity hybrids along with a 1S %%@
meson. Concentrating again 
on the hybrids with exotic $J^{PC}$, the modes of interest are:
\bea
\psi&\to & ^3\PPp_{2} ~1^1\uS_0	\label{mode3}\\
\chi_{0,1,2}&\to&^3\PPp_{0,2}~1^3\uS_1 	\label{mode5}
\eea
The relation \rf{rel2} implies that to first order in the hybrid wavefunction there is a correlation %%@
of scale between the production rates of the above modes and those of conventional mesons. This %%@
relationship can be exploited to predict the branching fractions for the above modes relative to %%@
observed conventional meson 
modes. In general there is not an immediate relationship between the full decay amplitudes for any of %%@
the above $\PPp+1\uS$ modes  
and their conventional meson counterparts $1\uS+1\S\uP$: charge conjugation requires that the modes %%@
have opposite spin so the 
angular momentum recouplings give different linear combinations of $L,l_f$ matrix elements of the type %%@
\rf{sme1} and \rf{sme2} in the full amplitude. 
The $\psi$ decay mode above is a fortunate exception. It has been shown that the decays $\psi\to %%@
b_1\pi$ and $\psi\to a_2\rho$ are 
consistent with the hypothesis that the virtual $q\q$ state through which the decay proceeds is a %%@
$\tso$ state. For $\psi\to a_2\rho$ in particular the decomposition is uniquely in terms of the %%@
spatial matrix element on the right hand side of equation \rf{rel2} (eqn. \rf{earlier}),
\be
\wbk{^3\S\uP_2~1^3\uS_1}{n^3\uS_1}=\frac{1}{2}\rme{((1\uP\ot 1\uS)_1\ot \uD)_{1}}{\grad}{n^3\uS_1}^2
\ee
while the analogous hybrid mode is expressed in terms of the spatial matrix element on the left hand %%@
side of equation \rf{rel2} \cite{tjbthesis}
\be
\wbk{^3\PPp_2~1^1\uS_0}{n^3\uS_1}=\frac{1}{8}\rme{((\PPp\ot 1\uS)_1\ot \uD)_{1}}{\grad}{n^3\uS_1}^2
\ee
Using eqn.\rf{rel2} there is a direct correlation of scale between the production amplitudes to first %%@
order in the hybrid 
wavefunction
\be
\wbk{^3\PPp_2~1^1\uS_0}{n^3\uS_1}\approx\frac{1}{8}\wbk{^3\S\uP_2~1^3\uS_1}{n^3\uS_1}\label{eight%%@
h}
\ee
%The scale associated with the string overlap $\kappa\sqrt{2\sigma}/\beta$ is close to unity and %has %%@
not been made explicit here. 

In the following the implications for the isovector $2^{+-}$ hybrid, denoted $b_2$, will be %%@
considered; predictions for the 
isoscalar hybrid follow after adjustments due to flavour and phase space. The momentum of the $b_2\pi$ %%@
mode is very close to that 
of $a_2\rho$ if the $b_2$ mass around  2 GeV as expected, so that the above eqn. \rf{eighth} %%@
translates into a direct relation between  
their branching fractions if the difference in the external phase space factors is disregarded. The %%@
$a_2\rho$ mode is second only to 
$\rho\pi$ in magnitude and so even with the above suppression by a factor of 8 the corresponding %%@
hybrid mode $b_2\pi$ should be 
observable:
\be
b.r.(\psi\to b_2\pi)\sim 1\times 10^{-3} \label{firstorder}
\ee
This result makes no reference to the initial state wavefunction other than that it is $^3\uS_1$. This %%@
first order estimate can only be taken as a guide as the leading order $\PPp$ wavefunction captures %%@
only the leading order angular dynamics.
% and has essentially the wrong radial wavefunction. 
Corrections with a more realistic wavefunction are discussed in ref. \cite{tjbthesis} and it appears %%@
that the above is an underestimate.

\section{Conclusion}

Within the assumption that \cc~ $\to $ \qq~ is the dominant intermediate state in the production
of light hadrons, and that the strong OZI decay amplitude for \qq~ $\to q\Q+Q\q$ 
factorises in the sense of ref.\cite{bct}, data imply that the intermediate \qq~ state is
not a hybrid meson. 

Although we have no well-developed dynamical model for this, it seems to us likely that this is %%@
consistent
with the general expectation that hybrid \qq~ production at short distances is
suppressed, due to the nodes in both the \qq~ and flux-tube wavefunctions. In turn this would imply %%@
that hybrid production from electromagnetic or weak currents, e.g. in $e^+e^-$ or $B$ decays will be %%@
enfeebled. We note that the $\psi(4260)$, which has characteristics of hybrid
charmonium\cite{fcpage,rumsfeld}, has a nugatory
leptonic width of O(eV), which was only exposed by study of the unusual channel $\psi \pi\pi$. 
If this state is not associated with hybrid charmonium, then even smaller leptonic widths
would need to be accessed. Models or lattice QCD are needed to give insight into
the short distance behaviour of hybrid wavefunctions and to assess whether they may couple
dominantly through intermediate $q\Q+Q\q$ loops. If the latter are important, this would go beyond
our analysis, wihch has restricted itself to \qq~ states.

The production of hybrid via the long-range components of its wavefunction,
such as OZI production in association with a conventional meson in \cc~ decays, depends on
the parity of the hybrid. Hybrids with negative parity are predicted to be suppressed;
hence if either of $\pi_1(1400/1600)$ is a hybrid meson, we do not anticipate a significant
signal in $\chi_1 \to \pi \pi_1$ for example. Conversely, if either is a $qq\bar{q}\bar{q}$ member of
a {\bf 10} or $\bar{\bf{10}}$, there is no selection rule against their production. Hence a search for %%@
$\chi_1 \to \pi \pi_1$ is merited; a strong signal
would be interesting in its own right, as well as being a possible indicator that
the exotic $1^{-+}$ $\pi_1$ signal is not a hybrid meson.

Our results suggest that positive parity hybrid production may be more promising. The
prediction that $\psi \to \pi b_2$ has a branching ratio of $\sim 10^{-3}$ would make this
typical in magnitude to other decays that have been studied successfully. The $b_2$ is expected to %%@
occur
in the 2 to 2.5 GeV region, where multibody decays may hinder its identification.
However, the mode $b_2 \to \pi a_2$ is predicted to be a dominant quasi two-body channel\cite{cp95}
and hence this signal may be extracted in $\psi \to 5 \pi$. We would recommend a high statistics study %%@
of $\psi$
 at BES in the hope of isolating this $J^{PC}$ exotic hybrid meson. 

This work is supported,
in part, by grants from
the Particle Physics and
Astronomy Research Council, the Oxford University
Clarendon Fund and the
EU-TMR program ``Eurodice'', HPRN-CT-2002-00311.

\bibliographystyle{h-physrev3}
\bibliography{tjb}
\end{document}